\documentclass[twoside,fleqn]{article}
\usepackage{espcrc2}
\usepackage{amssymb}
\usepackage[dvips]{epsfig}
\newcommand{\be}{\begin{equation}}
\newcommand{\ee}{\end{equation}}
\newcommand{\bea}{\begin{eqnarray}}
\newcommand{\eea}{\end{eqnarray}}
\newcommand{\bean}{\begin{eqnarray*}}
\newcommand{\eean}{\end{eqnarray*}}
\newcommand{\T}{\textstyle}

\newcommand{\ketbra}[3]{\langle#1|#2|#3\rangle}

\newcommand{\psib}{\psi_{\rm b}}
\newcommand{\bpsid}{\overline{\psi}_{\rm d}}

\newcommand{\lightb}{\overline{\psi}_{\rm l}}
\newcommand{\heavy}{\psi_{\rm h}}
\newcommand{\heavyb}{\overline{\psi}_{\rm h}}

\newcommand{\gfv}{\gamma_5}

\newcommand{\gbar}{\overline{g}}
\newcommand{\gbSF}{\overline{g}_{{\rm SF}}}
\newcommand{\gbsq}{\overline{g}^2}

\newcommand{\gbsqSF}{\overline{g}^2_{{\rm SF}}}

\newcommand{\MSbar}{\overline{\rm MS}}
\newcommand{\MS}{\MSbar}

\newcommand{\csw}{c_{\rm sw}}

\newcommand{\ct}{c_{\rm t}}
\newcommand{\rmd}{{\rm d}}

\newcommand{\lmax}{L_{\rm max}}
\newcommand{\Lmax}{L_{\rm max}}
\newcommand{\lmatch}{L_{\rm m}}
\newcommand{\mumatch}{\mu_{\rm m}}

\newcommand{\Or}{\mbox{O}}
\newcommand{\Fm}{{\rm fm}}
\newcommand{\MeV}{{\rm MeV}}

\newcommand{\half}{{\textstyle{\frac{1}{2}}}}
\newcommand{\lnab}[1]{{\overleftarrow{\nabla}\kern-1.5pt_{#1}}}
\newcommand{\lnabstar}[1]{\overleftarrow{\nabla}
            \kern-0.5pt\smash{\raise 4.5pt\hbox{$\ast$}}
            \kern-4.5pt_{#1}}
\newcommand{\zastat}{Z_{\rm A}^{\rm stat}}
\newcommand{\zastatSF}{Z_{\rm A,SF}^{\rm stat}}
\newcommand{\Astat}{A^{\rm stat}_0}
\newcommand{\Aistat}{(A^{\rm stat}_{\rm I})_{0}}
\newcommand{\Arstat}{(A_{\rm R}^{\rm stat})_{0}}
\newcommand{\sigastat}{\sigma_{\rm A}^{\rm stat}}
\newcommand{\Sigastat}{\Sigma_{\rm A}^{\rm stat}}
\newcommand{\castat}{c_{\rm A}^{\rm stat}}

\newcommand{\fastat}{f_{\rm A}^{\rm stat}}
\newcommand{\fone}{f_{1}}
\newcommand{\fonehh}{f_{1}^{\rm hh}}

\newcommand{\mbMSbar}{m_{{\rm b},\MS}}
\newcommand{\mB}{m_{\rm B}}
\newcommand{\Fb}{F_{\rm B}}
\newcommand{\Fbbare}{F_{\rm B}^{\rm bare}}
\newcommand{\Fbsstat}{F_{{\rm B}_{\rm s}}^{\rm stat}}
\newcommand{\FhatSF}{\hat{F}_{\rm SF}}
\newcommand{\Fhatstat}{{\Phi}^{\rm stat}}
\newcommand{\Phibare}{\Phi_{\rm bare}}
\newcommand{\PhiRGI}{\Phi_{\rm RGI}}

\newcommand{\Phimatch}{\Phi_{\rm match}}
\newcommand{\ZPhi}{Z_{\Phi}}
\title{
{
\vspace{-4.25cm} \normalsize \hfill
\parbox{25.0mm}{\raggedleft
MS-TP-02-3\\SHEP 02-20\\DESY 02-130\\September 2002}
}\\[25mm]
Non-perturbative determination of $\zastat$ in quenched QCD%
\thanks{Based on a talk presented by J.H. at the conference
LATTICE '02, June 24 -- 29, 2002, MIT, Cambridge MA, USA.}%
\thanks{Supported by the EU under HPRN-CT-2000-00145.}
}
\author{
Jochen Heitger\address{
Universit\"at M\"unster, Institut f\"ur Theoretische Physik,
D-48149 M\"unster, Germany},
Martin Kurth\address{
University of Southampton, Department of Physics and Astronomy, 
Southampton SO17~1BJ, UK} and
Rainer Sommer\address{
Deutsches Elektronen-Synchrotron DESY, D-15738 Zeuthen, Germany}
(ALPHA Collaboration)
}
\begin{document}
%
%%%%%%%%%%%%%%%%%%%%%%%%%%%%%%%%%%%%%%%%%%%%%%%%%%%%%%%
% slight change in table style (J.H.,1996) %%%%%%%%%%%%
\makeatletter
\long\def\@maketablecaption#1#2{\vskip 10mm #1. #2\par}
\makeatother
%%%%%%%%%%%%%%%%%%%%%%%%%%%%%%%%%%%%%%%%%%%%%%%%%%%%%%%
%
\begin{abstract}
We non-perturbatively calculate the renormalization factor of the static 
axial vector current in $\Or(a)$ improved quenched lattice QCD.
Its scale dependence is mapped out in the Schr\"odinger functional 
scheme by means of a recursive finite-size scaling technique, taking the 
continuum limit in each step.
We also obtain $\zastat$ for Wilson fermions in order to renormalize 
existing unimproved data on $\Fbbare$ non-perturbatively.
\end{abstract}
\maketitle
%
%%%%%%%%%%%%%%%%%%%%%%%%%%%%%%%%%%%%%%%%%%%%%%%%%%%%%%%%%%%%%%%%%%%%%%%%%
%
\section{INTRODUCTION}
The decay constant $\Fb$, governing the leptonic decay of a B-meson, is
a quantity of much phenomenological interest.
However, as heavy flavours on the lattice escape a direct numerical 
treatment, the \emph{static approximation} represents a valuable 
alternative, since this effective theory has simplified dynamics and 
describes the large mass limit of the theory.
Yet the problems of this framework have been twofold in the past.
Owing to the infinitely heavy b-quark, (i) the renormalization properties 
of the static theory are different, i.e.~the renormalization constant 
$\zastat$ of the axial current in
$\Arstat=\zastat(\mu)\,\bpsid\gamma_0\gfv\psib^{\rm stat}$ becomes scale
($\mu$) dependent, thereby entailing an additional uncertainty, and 
(ii) MC computations of the matrix element itself are difficult.

Here, we solve (i) by \emph{matching through the renormalization group 
invariant (RGI) operator}:
The non-perturbatively calculated $\mu$--dependence of any renormalized 
matrix element of $\Astat$, 
$\Phi\equiv\Fhatstat=\ketbra{\,{\rm f}\,}{\Arstat}{\,{\rm i}\,}$,
together with the value of $\zastat$ at some (low) $\mu$, yields a factor 
leading to the RGI $\Phi$.
It is then related to the `matching' scheme, which approximates the 
relativistic theory up to $1/m$--corrections.
For a more detailed explanation of the overall strategy we refer 
to \cite{lat02:rainer} and for a full report on our work 
to \cite{zastat:pap3_prep}.
\section{STRATEGY}
\begin{figure}[htb]
\centering
\vspace{-2.0cm}
\epsfig{file=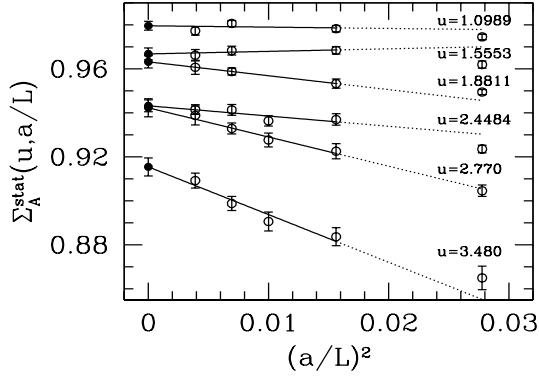,width=7.5cm,height=7.0cm}
\vspace{-2.25cm}
\caption{{\sl
Lattice step scaling function and its continuum extrapolation for some 
selected $u$.
}}\label{fig:Sigma_zastat1}
\vspace{-1.0cm}
\end{figure}

Our determination of $\zastat$ and its running with $\mu$ uses the 
Schr\"odinger functional (SF) \cite{SF:LNWW} as intermediate 
scheme \cite{impr:lett}.
The procedure resembles ALPHA's quark mass 
renormalization \cite{msbar:pap1}:\\[0.8ex]
1.~Choose a proper combination $X$ of SF correlation functions such that, 
   with $X_{\rm R}=\zastat X$ finite and
   $X=X^{(0)}+X^{(1)}g_0^2\,+\,\cdots\,$, the condition
   $X_{\rm R}\equiv X^{(0)}$ non-perturbatively defines $\zastat$,
   running with the SF's box size $L=1/\mu$.\\[0.8ex]
2.~Map out the $L$--dependence recursively by employing the finite-size 
   step scaling function (SSF)
   $\sigastat(u)=\lim_{a\rightarrow 0}\,\Sigastat(u,a/L)$,
   \be
   \Sigastat\left(u,{\T \frac{a}{L}}\right)\equiv
   \frac{\zastat(g_0,2L/a)}{\zastat(g_0,L/a)}
   \,,\,\,\,
   u\equiv\gbsqSF(L)\,,
   \ee
   computed on the lattice and extrapolated to the continuum as 
   illustrated in fig.~\ref{fig:Sigma_zastat1}.\\[0.8ex] 
3.~Evolve $\zastat$ from initially large $L$ (low $\mu)$ to small $L$ 
   (high $\mu$) by repeatedly applying $[\sigastat]^{-1}$; 
   from there, continue with the perturbative high-energy behaviour to 
   arrive at (scale and scheme independent) RGI matrix elements
   \bea
   \PhiRGI
   & = &
   \Phi(\mu)\times\left(2b_0\gbar^2\right)^{-\gamma_0/2b_0} 
   \nonumber\\
   &   &
   \times\exp\bigg\{-\int_0^{\gbar} \rmd g\,\Big[\,
   {\gamma(g)\over\beta(g)}-{\gamma_0 \over b_0 g}\,\Big]\bigg\}
   \label{me_RGI}
   \eea
   of $\Arstat$, with $\gbar=\gbSF(L=1/\mu)$ and universal coefficients 
   $b_0=11/(4\pi)^2$ and $\gamma_0=-1/(4\pi^2)$.\\[0.8ex]
Finally, for sufficiently high $\mu$, the RG invariant $\PhiRGI$ can be 
converted to the `matching' scheme via a perturbative factor 
$\Phimatch(\mu)/\PhiRGI$.
%
%%%%%% figure: Sigma_A^stat ("new" scheme only)
%
\section{SURVEY OF THE RESULTS}
The simulation and analysis as well as the SF setup are basically 
analogous to \cite{msbar:pap1}, except that the boundary coefficient 
$\ct$ is set to 2--loop \cite{impr:ct_2loop}.
While the inclusion of static quarks and their impact on $\Or(a)$
improvement are described in \cite{zastat:pap1}, we only recall the 
heavy quark lattice action \cite{stat:eichhill1},
\be
S_{\rm h}[U,\heavyb,\heavy]= 
a^4\sum_x\heavyb(x)\nabla_{0}^{\ast}\heavy(x)
\ee
($\nabla_{0}^{\ast}$ the backward derivative's time component,
i.e.~static quarks propagate only forward in time), and the $\Or(a)$ 
improvement term of $\Astat$ in
\be
\Aistat=
\Astat+a\castat\lightb\gamma_j\gamma_5
\half(\lnab{j}+\lnabstar{j})\heavy,
\ee
where the 1--loop $\castat=-1/(4\pi)$ \cite{stat:ca} is used.

Although in ref.~\cite{zastat:pap1} a suitable renormalization condition 
for $\Astat$ in the SF scheme has already been formulated, we slightly 
modify it in the present non-perturbative context for reasons of 
numerical precision.
We impose instead
\be
X(u,a/L)\equiv
\frac{\fastat(L/2)}{\left[\,f_1\times\fonehh(L/2)\,\right]^{1/4}}
\,\bigg|_{\,\gbsq=u}\,,
\label{rencond}
\ee
at vanishing quark mass.
$\fastat$ is a correlator between a static-light pseudoscalar boundary 
source and $\Astat$ in the bulk, $\fone$ between two light-quark boundary 
sources, and $\fonehh$ denotes a correlator in $x_3$--direction of Wilson 
lines between static-light boundary sources, which is efficiently 
evaluated in MC by multi-hit \cite{zastat:pap3_prep}.
Note that the boundary wave function renormalizations and a linearly
divergent mass counterterm cancel in eq.~(\ref{rencond}).

\begin{figure}[htb]
\centering
\vspace{-1.75cm}
\epsfig{file=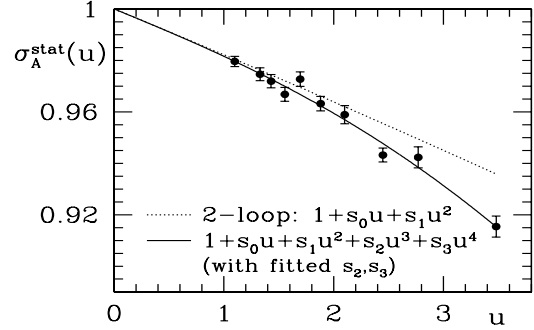,width=7.5cm,height=6.5cm}
\vspace{-2.625cm}
\caption{{\sl
Continuum SSF and polynomial fit.
}}\label{fig:sigma_zastat1}
\vspace{-0.645cm}
\end{figure}

\begin{figure}[htb]
\centering
\vspace{-1.75cm}
\epsfig{file=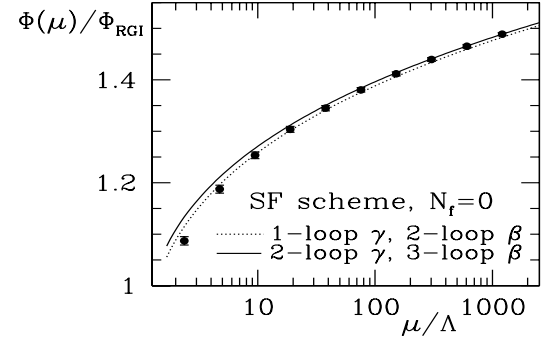,width=7.5cm,height=6.5cm}
\vspace{-2.375cm}
\caption{{\sl
Running matrix element of $\Arstat$ in the SF scheme compared to 
perturbation theory.
}}\label{fig:PhiSF_stat}
\vspace{-0.75cm}
\end{figure}

Fig.~\ref{fig:sigma_zastat1} displays the continuum SSF after 
$a\rightarrow 0$ extrapolation of $\Sigastat$ together with an 
interpolating fit.
%
%%%%%% figure: sigma_A^stat ("new" scheme only)
%
$\sigastat$ being known, it connects $\Phi(\mu)$ in the low-energy regime 
at $\mumatch=1/\lmatch\equiv(2\lmax)^{-1}$ step-by-step with the 
perturbative domain, and once $\mu$ is large enough for perturbative 
running to set in, we find by numerical integration 
in eq.~(\ref{me_RGI}):
\be
\Phi(\mu)/\PhiRGI=1.088(8)
\quad {\rm at} \quad \mu=\mumatch\,.
\ee
In fig.~\ref{fig:PhiSF_stat} we compare the numerically computed  
running of the static axial current in the SF scheme with perturbation 
theory, where also $\Lambda\Lmax=0.211$ from ref.~\cite{msbar:pap1} 
enters the analysis.
%
%%%%%% figure: Phi_RGI^stat/Phi^stat(mu) (SF scheme)
%
\subsection{The total renormalization factor}
\begin{figure}[htb]
\centering
\vspace{-1.75cm}
\epsfig{file=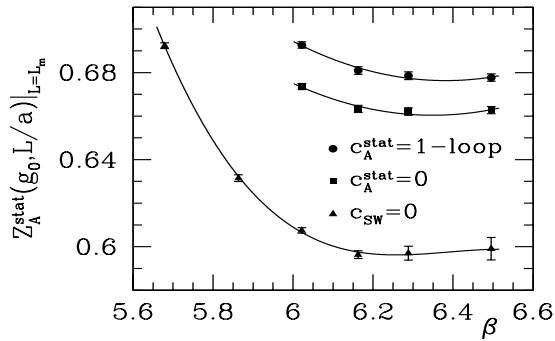,width=7.5cm,height=6.5cm}
\vspace{-2.375cm}
\caption{{\sl
Results for $\zastat$ at scale $L=1.436\,r_0$.
}}\label{fig:zastat1_1.436r0}
\vspace{-0.625cm}
\end{figure}

We still need to relate $\Arstat(\mu)$ to the bare lattice operator.
This amounts to compute $\zastat$ at the (low-energy) matching scale
$\lmatch=1.436\,r_0$.
Fig.~\ref{fig:zastat1_1.436r0} shows numerical results for 
$\zastat(g_0,L/a)|_{L=\lmatch}$ and polynomial fits, where we also 
studied the cases $\castat=0$ and $\csw=0$ for later purposes.
%
%%%%%% figure: Z_A^stat(g_0,L/a)_{L=2*L_max}
%
The total renormalization factor to translate any bare matrix element 
$\Phibare(g_0)$ of $\Astat$ into the RGI one, $\PhiRGI$, then reads
\be
\ZPhi(g_0)=
\frac{\PhiRGI}{\Phi(\mu)}\,\bigg|_{\,\mu=\mumatch}\times
\zastat(g_0,L/a)\,\Big|_{\,L=\lmatch}
\ee
with $\PhiRGI/\Phi(\mumatch)$ a universal part independent of the
regularization and $\zastat(g_0,L/a)|_{L=\lmatch}$ that depends on it.
$\ZPhi$ will be given explicitly in \cite{zastat:pap3_prep}.
\subsection{Wilson data on $\Fb$ revisited}
Since we now also have the SF renormalization factor 
$\zastat(g_0,L/a)|_{L=\lmatch}$ for unimproved Wilson fermions
($\csw=0$) available, it may be combined with the universal part 
$\PhiRGI/\Phi(\mu)$ at $\mu=\mumatch$ and the perturbative conversion 
factor $\Phimatch(\mu)/\PhiRGI$ at scale $\mu=\mbMSbar$ in order to 
confront it with tadpole-improved estimates of the FNAL 
group \cite{stat:fnal2} on the $Z$--factor, which equals 
$\Phimatch(\mbMSbar)/\Phibare(g_0)$ in our notation.
As it turns out (cf.~\cite{lat02:rainer}), they deviate significantly 
in the relevant range of $g_0$, which reveals the importance of 
\emph{non-perturbative} renormalization.

\begin{figure}[htb]
\centering
\vspace{-1.375cm}
\epsfig{file=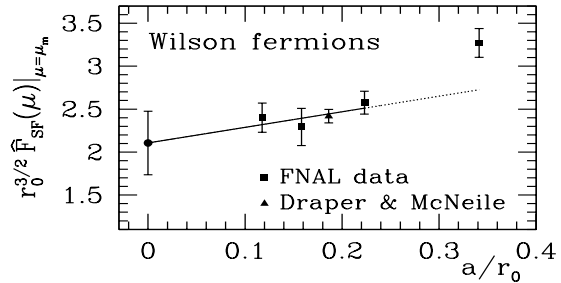,width=7.5cm,height=6.5cm}
\vspace{-2.75cm}
\caption{{\sl
Continuum extrapolation of the non-perturbatively renormalized
$\Fbbare$ from \cite{stat:fnal2,stat:DMcN94}.
}}\label{fig:fBr1.436r0_wil}
\vspace{-0.875cm}
\end{figure}

Just so, we non-perturbatively renormalize existing Wilson data
\cite{stat:fnal2,stat:DMcN94} on $\Phibare=\Fbbare\sqrt{\mB}$,
\be
\zastatSF\,\big|_{\,\lmatch}\times\,
r_0^{3/2}\,\Phibare
\equiv
r_0^{3/2}\,\FhatSF\,\big|_{\,\mumatch}\,,
\ee
and extrapolate them to $a=0$ in fig.~\ref{fig:fBr1.436r0_wil} discarding
the rightmost point ($\beta=5.7$).
With $r_0=0.5\,\Fm$, this results in $\Fbsstat=261(46)\,\MeV$ at
$\mu=\mbMSbar$ in the `matching' scheme, containing all errors
apart from quenching.  
%
%%%%%% figure: F_B for Wilson fermions
%
\section{CONCLUSIONS}
We have performed the scale dependent renormalization of $\Astat$ by
constructing a non-perturbative RG in the SF scheme, and agreement with 
perturbation theory at large scales is demonstrated.
The renormalization factors needed to get the associated RG invariants 
are computed with good numerical accuracy \cite{zastat:pap3_prep}, which 
is a crucial prerequisite for a controlled determination of $\Fb$ in the 
static limit.

Our continuum extrapolation that uses unimproved data from the 
literature and also quite large lattice spacings leaves much room to
improve on the present result $\Fbsstat=261(46)\,\MeV$.
%
%%%%%%%%%%%%%%%%%%%%%%%%%%%%%%%%%%%%%%%%%%%%%%%%%%%%%%%%%%%%%%%%%%%%%%%%%
%
% bibliography
%

%

\begin{thebibliography}{99}
%
\bibitem{lat02:rainer}
R. Sommer, these proceedings.
%
\bibitem{zastat:pap3_prep}
J. Heitger, M. Kurth and R. Sommer,
in preparation.
%
\bibitem{SF:LNWW}
M. L\"uscher, R. Narayanan, P. Weisz and U. Wolff,
Nucl. Phys. B384 (1992) 168.
%
\bibitem{impr:lett}
K. Jansen et al.,
Phys. Lett. B372 (1996) 275.
%
\bibitem{msbar:pap1} 
S. Capitani, M. L\"uscher, R. Sommer and H. Wittig, 
Nucl. Phys. B544 (1999) 669.
%
\bibitem{impr:ct_2loop}
A. Bode, P. Weisz and U. Wolff, 
Nucl. Phys. B576 (2000) 517.
%
\bibitem{zastat:pap1}
M. Kurth and R. Sommer,
Nucl. Phys. B597 (2001) 488.
%
\bibitem{stat:eichhill1}
E. Eichten and B. Hill,
Phys. Lett. B234 (1990) 511.
%
\bibitem{stat:ca}
C. Morningstar and J. Shigemitsu,
Phys. Rev. D57 (1998) 6741.
%
\bibitem{stat:fnal2}
A. Duncan et al.,
Phys. Rev. D51 (1995) 5101.
%
\bibitem{stat:DMcN94}
T. Draper and C. McNeile,
Nucl. Phys. Proc. Suppl. 34 (1994) 453.
%
\end{thebibliography}
\end{document}